\documentclass[twoside,12pt]{article}  %%% ���� LaTeX ��ʽ�ļ�������ָ�����ȱ��
\usepackage{CJK}   %%%%% ʹ�����ġ����ġ�����ʱ���ô˺���
\usepackage{indentfirst}  %% �ڱ�������һ�ж�����
\usepackage{bm}    %%%%%  ��ӡ��ѧ���ź�б��ʱ���ô˺���
\usepackage{amsmath}
\usepackage{color}
\usepackage{graphicx}
\usepackage{subfigure}
\usepackage{amsmath,amsfonts,amssymb,graphicx}
\usepackage{appendix}

\usepackage{graphicx}  %%%% ���в��� eps ͼ��ʱ���ô˺�������ͼ���� 1
\begin{document}    %% �ı��ļ���ʼ�����Ǳ�����ָ��

\begin{CJK*}{GBK}{song}  %% ��ʼ�������Ļ���

\begin{center}
\LARGE\bf  Fluid reactive anomalous transport  with random waiting time depending on the preceding jump length$^{*}$   %% ������Ŀ
\end{center}

\footnotetext{\hspace*{-.45cm}\footnotesize $^*$Project supported by the National Natural Science Foundation of China (Grant
No. 11626047) and the Foundation for Young Key Teachers of Chengdu
University of Technology, China (Grant No. KYGG201414).} \footnotetext{\hspace*{-.45cm}\footnotesize
$^\dag$Corresponding author. E-mail: zhanghong13@cdut.cn (Hong Zhang), liguohua13@cdut.cn (Guo-Hua Li)}

\begin{center}
\rm Hong Zhang, \ \ Guo-Hua Li
\end{center}

\begin{center}
\begin{footnotesize} \sl
%${}^{\rm a)}$ \\   %%%% ��ַ a)
Department of Mathematics Teaching, Chengdu University of
Technology, Cheng'du, Si'chuan 610059, China
%${}^{\rm b)}$ \\   %%%% ��ַ b)
%${}^{\rm c)}$ \\   %%%% ��ַ c)
%%% ������ַ������������
\end{footnotesize}
\end{center}

\begin{center}
\footnotesize (Received xx xx 2018; revised manuscript received xx
xx 2018)
          %% (Received �� �� ��; revised manuscript received �� �� ��)
\end{center}

\vspace*{2mm}

\begin{center}
\begin{minipage}{15.5cm}
\parindent 20pt\footnotesize

Anomalous (or non-Fickian) diffusion has been widely found in fluid reactive transport and the traditional advection diffusion reaction equation based on Fickian diffusion is proved to be
 inadequate to predict this anomalous transport of the reactive particle in flows.
To capture the complex couple effect among advection, diffusion and reaction, and the energy-dependent characteristics of fluid reactive anomalous transport, in the present paper we analyze $A\rightarrow B$ reaction under anomalous diffusion with waiting time depending on the
preceding jump length in linear flows, and derive the corresponding master equations in Fourier-Laplace space for the distribution of A and B particles  in continuous time random walks scheme. As examples, the generalized
advection diffusion reaction equations for the jump length of Gaussian distribution and l$\acute{e}$vy flight with
the probability density function of waiting time being quadratic dependent on the preceding jump length are obtained by
applying the derived master equations.

\end{minipage}

\end{center}

\begin{center}
\begin{minipage}{15.5cm}
\begin{minipage}[t]{2.3cm}{\bf Keywords:}\end{minipage}
\begin{minipage}[t]{13.1cm}
%%%%% �ؼ���
 Anomalous (or non-Fickian) diffusion, Continuous time random walk, Advection diffusion reaction equation

\end{minipage}\par\vglue8pt
{\bf PACS: }
%%% PACS ������
%% ��ѯ��ַ��http://www.aip.org/pacs
 05.60.-k, 82.40.-g, 47.70.Fw, 05.90.+m
\end{minipage}
\end{center}

\section{Introduction}  %%% �ڱ��� 1

Fluid reactive transport is an important field of
study in hydrogeology and other sciences that has a variety of
applications such as the transport of contaminants in
underground water$^{[1]}$, nuclear waste storage$^{[2]}$,  and carbon dioxide
(CO$_2$) storage$^{[2]}$, etc.
The macroscopic description of  coupled reaction and
transport is the standard advection diffusion
reaction equation (ADRE) based on Fickian diffusion in one-dimensional form as:
\begin {equation}\label{ADRE}
\frac{\partial C(x,t)}{\partial t}+v\frac{\partial C(x,t)}{\partial
x}=K\frac{\partial^2 C(x,t)}{\partial x^2}+f
\end {equation}
where $C(x,t)$ is the  probability density function (PDF) of the
particle, $v$ is constant velocity, $K$ is diffusion coefficient, and $f$ denotes the decoupled reaction term.

However, in recent years many tracer tests in natural complex porous media are found to exhibit
anomalous diffusion (non-Fickian) behavior deviating from Fickian diffusion,and even without reaction ($f=0$ in Eq.(1))  the
classical advection-dispersion equation(ADE) has proven to
be unsuitable for describing this kind of tracer transport.$^{[2-6]}$
One of the effective ways to quantify anomalous (or non-Fickian)
transport in nonhomogeneous porous media is power law waiting time continuous time random
walks (CTRW$'$s) model.$^{[7-9]}$ Based on various CTRW models
scholars obtain several generalization types of ADE which can be
used to describe the evolution of the probability density function
(PDF) of the particles undergoing anomalous dispersion in flow
fields.$^{[10-11]}$

On the other hand, there are some reaction-anomalous diffusion equations  where the reaction term has nontrivial coupled effect on diffusion term proposed.$^{[12-16]}$ More recently
a modified fractional ADRE,
\begin {equation}\label{FADRE}
\frac{\partial C_{m}(x,t)}{\partial t}+\frac{\partial^{r} C_{m}(x,t)}{\partial
t^{r}}=K\frac{\partial^{2} C_{m}(x,t)}{\partial x^{2}}-\frac{t^{-r}}{\Gamma(1-r)}C_{m,0}
\end {equation}
was derived  for the transport of mobile concentration of upscaling chemical reactions in multicontinuum systems.$^{[17]}$  Eq.(2) convolves a memory function $\frac{t^{-r}}{\Gamma(1-r)}$ and the time derivative term, and shows  that the description of the upscaling procedure in
the presence of reactions  is not a fractional diffusion equation with a naive reaction term
added. It should be noted that for the sake of simplicity in the above transport
equation the drift (advection) term is not included , and is considered to be relatively straightforward. But in fact, the advection term maybe have a complex couple effect on the anomalous diffusion and the chemical reaction terms in some circumstances$^{[9,18]}$, and should be further considered.

In 2006, Zaburdaev proposed a generalized CTRW model for certain circumstances in which the waiting time
is connected with the preceding jump length. This model describes
the dependence of the random waiting time on the energy and suggests
a method that includes the details of the microscopic distribution
over the waiting times and arrival distances at a given
point.$^{[19]}$ Note that up to now this coupled CTRW model
has been not associated with the fluid reactive transport. In the present paper we shall consider the simple A$\rightarrow$B reaction under anomalous diffusion
on moving fluid based on this CTRW model
with random waiting time depending on the preceding jump length, which can not only show the coupling among advection velocity, anomalous diffusion and chemical reaction,
but also describe the
energy-dependent characteristics of fluid reactive anomalous transport.
As examples, we shall apply the
new master equation to derive four generalized ADRE$'$s for Gaussian
distribution and l$\acute{e}$vy flight for jump lengths in linear flows when the
microscopic distribution of the random sojourn time is quadratic
dependent on the preceding jump length.

\section{Coupled CTRW model with random waiting time depending on the preceding jump length}

We first recall the coupled CTRW model with waiting time depending on the
spent energy or the preceding jump length  in one-dimensional lattices proposed by
Zaburdaev.$^{[19]}$ In this model, each step of the particle
requires some energy, and after making a jump a particle needs time
to recover. The longer the preceding jump distance, the longer are
the recovery and the waiting time. It means that the PDF of the
waiting time $\psi(|y|,t)$ before making the second step depends
both on the length of the preceding jump $|y|$ and the waiting time
$t$. Thus, the particle jumps from $x-y$ to $x$ with the jump length
PDF $\lambda(y)$, and then waits at $x$ for time $t$ drawn from
$\psi(|y|,t)$, after which the process is renewed. By assuming that
in the initial state all particles have zero arrival distances and
zero resting times, one obtained the balance equation for the PDF
$\rho(x,t)$ of the particles
\begin{eqnarray}
\rho(x,t)&=&\int_{-\infty}^{+\infty}dx'\int_{0}^{t}j(x',t')\lambda(x-x')\Psi(|x-x'|,t-t')dt'+\Psi_{0}(t)\rho_{0}(x).
\end{eqnarray}
Here, $j(x,t)$ is the escape rate, and satisfies the equation
\begin{eqnarray}
j(x,t)&=&\int_{-\infty}^{+\infty}dx'\int_{0}^{t}j(x',t')\lambda(x-x')\psi(|x-x'|,t-t')dt'+\psi_{0}(t)\rho_{0}(x),
\end{eqnarray}
where the survival time distribution
$\Psi(|y|,t)=1-\int_{0}^{t}\psi(|y|,\tau)d\tau$ depends both on the
waiting time and the preceding jump length, the term
$\Psi_{0}(t)\rho_{0}(x)=\Psi(|0|,t)\rho(x,0)$ is supposed as the
influence of the initial distribution, and $\psi_{0}(t)=\psi(|0|,t)$
is the waiting time PDF for the initial position. In Fourier-Laplace
space, the PDF $\rho(x,t)$ obeys the master equation for the CTRW
model with waiting time depending on the preceding jump length:
\begin{equation}\label{Zmaster}
\rho(k,u)=\Psi_{0}(u)\rho_{0}(k)+\frac{\{\Psi(|x|,u)\lambda(x)\}_{k}\psi_{0}(u)\rho_{0}(k)}{1-\{\psi(|x|,u)\lambda(x)\}_{k}},
\end{equation}
where  the symbol $\psi_{0}(u)$ is the Laplace transform of
$\psi_{0}(t)$, and $\Psi_{0}(u)\rho_{0}(k)$,
$\{\Psi(|x|,u)\lambda(x)\}_{k}$, $\{\psi(|x|,u)\lambda(x)\}_{k}$
denote the Fourier-Laplace transform of $\Psi_{0}(t)\rho_{0}(x)$,
$\Psi(|x|,t)\lambda(x)$, $\psi(|x|,t)\lambda(x)$, respectively.

\section{Fluid reactive anomalous transport  with random waiting time depending on the preceding jump length}

We shall now consider
 the energy dependent random walk model in a moving fluid with an inhomogeneous velocity field $v(x)$.
Noting that in Ref.[9], [10] and [20] in Galilei
variant model the jump length  $y$ for the moving particle dragged
along the velocity $v(x)$ is replaced by $ y-\tau_{a}v(x)$, where
$\tau_{a}$ stands for an advection time scale, and $\tau_{a}v(x)$ is
the mean drag experienced by a particle jumping from the point $x$,
 in our
model we introduce the PDF of a length of step $y$
with waiting time $t$ which will depend on the velocity $v(x)$ of
the fluid with the starting point $x$ of the jump, i.e.,
$\lambda(y-\tau_{a}v(x))\Psi(|y-\tau_{a}v(x)|,t)$.
 When $v(x)=0$,
this PDF recovers the coupled density
$\lambda(y)\Psi(|y|,t)$ for the particle in Ref. [19].

We then study the simplest reaction scheme $A\rightarrow B$ in
this CTRW model. We assume all physical properties of A and B particles are
the same and the particles trapped in stagnant regions will react
with a relabeling of A into B taking place at a rate $\alpha$ without changing energy. Let
$A(x,t)$ be the PDF of A particle being in point $x$ at time $t$ and
and $i^{-}(x,t)$ be the escape rate. By assuming that in the initial
distribution all particles have zero resting times, we can find the
balance equation for A particles in a given point:
\begin{eqnarray}\label{A(x,t)}
A(x,t)&=&A_{0}(x)\Psi_{0}(t)e^{-\alpha
t}+\int_{-\infty}^{+\infty}dx'\int_{0}^{t}i^{-}(x',t')\lambda(x-x'-\tau_{a}v(x'))\nonumber\\
& &\times\Psi(|x-x'-\tau_{a}v(x')|,t-t')e^{-\alpha(t-t')}dt'
\end{eqnarray}
where $A_0(x)$ is  the initial state  of A particle,
$\Psi(|y|,t)e^{-\alpha t}=(1-\int_{0}^{t}\psi(|y|,\tau)d\tau)e^{-\alpha t}$
is the joint survival density of remaining at least at time $t$ on
the spot (without being converted into B). The density is a sum of
outgoing particles from all other points at different times given by
the flow, and provided they survived after their arrival till the
time $t$. The first term on the right hand side is just the
influence of the initial distribution.

Fourier transforming $x\rightarrow k$ and Laplace transforming
$t\rightarrow u$ of Eq.(\ref{A(x,t)}), we find
\begin{eqnarray}\label{A(k,u)}
A(k,u)&=&A_{0}(k)\Psi_{0}(u+\alpha)+[\Psi(i\frac{\partial}{\partial k},u+\alpha)\lambda(k)]\int_{-\infty}^{+\infty}i^{-}(x',u)e^{-k(x'+\tau_{a}v(x'))}dx'.
\end{eqnarray}
 In the above expression $A_0(k)$ represents the Fourier
$x\rightarrow k$ transform  of the initial condition $A_0(x)$,
$\Psi_{0}(u+\alpha)$ denotes the Laplace transform of joint survival PDF
$\Psi_{0}(t)e^{-\alpha t}$, $A(k,u)$ and $i^{-}(k,u)$ are the Fourier-Laplace
transforms of $A(x,t)$ and $i^{-}(x,t)$ respectively, and the term $\Psi(i\frac{\partial}{\partial k},u+\alpha)\lambda(k)$ represents the Fourier-Laplace
 transform of $\Psi(|x|,t)\lambda(x)$ where we use the property of Fourier transform $\mathcal{F}(xf(x))=i\frac{\partial}{\partial k}f(k)$.

To get the master equation with respect to $A(x,t)$, we shall give the other balance equation.
We notice that
that the loss flux is from those particles that were originally at
$x$ at $t =0$ and wait without reacting until time $t$ to leave, and
those particles that arrived at an earlier time $t'$ and wait
without reacting until time $t$ to leave, and have the second
balance equation:
\begin{eqnarray}\label{i-(x,t)}
i^{-}(x,t)&=&A_{0}(x)\psi_{0}(t)e^{-\alpha
t}+\int_{-\infty}^{+\infty}dx'\int_{0}^{t}i^{-}(x',t')
\lambda(x-x'-\tau_{a}v(x'))\nonumber\\
&&\times\psi(|x-x'-\tau_{a}v(x')|,t-t')e^{-\alpha(t-t')}dt'
\end{eqnarray}
where $\psi(|x-x'|,t-t')e^{-\alpha(t-t')}$ is the non-proper waiting time density
depending on the preceding jump length for the actually made new step provided the particle survived.
 By applying the transform $(x,t)\rightarrow (k,u)$ of Eq.(\ref{i-(x,t)}), we find
\begin{eqnarray}\label{i-(k,u)}
i^{-}(k,u)=A_{0}(k)\psi_{0}(u+\alpha)+
[\psi(i\frac{\partial}{\partial k},u+\alpha)\lambda(k)]\int_{-\infty}^{+\infty}i^{-}(x',u)e^{-k(x'+\tau_{a}v(x'))}dx'
\end{eqnarray}
where the term $\psi(i\frac{\partial}{\partial k},u+\alpha)\lambda(k)$ is the Fourier-Laplace
 transform of $\Psi(|x|,t)\lambda(x)$.
 Using (\ref{A(k,u)}) and (\ref{i-(k,u)}), one has
\begin{eqnarray}\label{Brelation}
\frac{A(k,u)-A_{0}(k)\Psi_{0}(u+\alpha)}{\Psi(i\frac{\partial}{\partial k},u+\alpha)\lambda(k)}=\frac{i^{-}(k,u)-A_{0}(k)\psi_{0}(u+\alpha)}{\psi(i\frac{\partial}{\partial k},u+\alpha)\lambda(k)},
\end{eqnarray}
from which we find
\begin{equation}\label{i-(k,u)=}
i^{-}(k,u)=A_{0}(k)\psi_{0}(u+\alpha)+\frac{\psi(i\frac{\partial}{\partial k},u+\alpha)\lambda(k)}{\Psi(i\frac{\partial}{\partial k},u+\alpha)\lambda(k)}[A(k,u)-A_{0}(k)\Psi_{0}(u+\alpha)].
\end{equation}

We assume a linear velocity $v(x)=\omega x$ where
$\omega$ is a constant. Then Eq.(\ref{A(k,u)}) becomes
\begin{equation}\label{A(k,u)v}
A(k,u)=\Psi(u+\alpha)A_{0}(k)+\Psi(i\frac{\partial}{\partial k},u+\alpha)\lambda(k)j(k+v_{k},u)
\end{equation}
where the symbol $v_{k}=\tau_{a}\omega k$.
 In the limit $\tau_{a}\rightarrow 0$, Eq.(\ref{A(k,u)v}) gives
 \begin{eqnarray}\label{A(k,u)vk}
A(k,u)&\simeq&\Psi(u+\alpha)A_{0}(k)+\Psi(i\frac{\partial}{\partial k},u+\alpha)\lambda(k)\nonumber\\
& &\times[i^{-}(k,u)+v_{k}i^{-}{'}_{k}(k,u)].
\end{eqnarray}
We substitute (\ref{i-(k,u)=}) into (\ref{A(k,u)vk})  and finally obtain the generalized master equation
in Fourier-Laplace space for the distribution of A-paricles  in $A\rightarrow B$ reaction-anomalous diffusion in fluid fields with waiting time depending the preceding jump length:
\begin{equation} \label{Amaster}
\mathcal{D}(k,u)A(k,u)=\mathcal{V}(k,u)A'_k(k,u)+\mathcal{I}(k,u)
\end{equation}
where
$$\mathcal{D}(k,u)=[1-\psi(i\frac{\partial}{\partial k},u+\alpha)\lambda(k)-v_{k}\Phi_{\alpha}(k,u)'_{k}\Psi(i\frac{\partial}{\partial
k},u+\alpha)\lambda(k)], ~~~~~~~~~~~~~~~~~~~~~~~~~$$$$
\mathcal{V}(k,u)=v_{k}[\psi(i\frac{\partial}{\partial
k},u+\alpha)\lambda(k)],~~~~~~~~~~~~~~~~~~~~~~~~~~~~~~~~~~~~~~~~~~~~~~~~~~~~~~~~~~~~~~~~~~~~~~~$$
$$\mathcal{I}(k,u)=-v_{k}[\Psi(i\frac{\partial}{\partial
k},u+\alpha)\lambda(k)]\Phi_{\alpha}(k,u)'_{k}\Psi_{0}(u+\alpha)A_{0}(k)~~~~~~~~~~~~~~~~~~~~~~~~~~~~~~~~~~~~~$$
$$~~~~~~~~~~~~+[1-\psi(i\frac{\partial}{\partial
k},u+\alpha)\lambda(k)]\Psi_{0}(u+\alpha)A_{0}(k)
+v_{k}[\Psi(i\frac{\partial}{\partial
k},u+\alpha)\lambda(k)]\psi_{0}(u+\alpha)A_{0}'(k)$$
$$~~~~~~~~~~~-v_{k}[\psi(i\frac{\partial}{\partial
k},u+\alpha)\lambda(k)]\Psi_{0}(u+\alpha)A_{0}'(k)
+[\Psi(i\frac{\partial}{\partial
k},u+\alpha)\lambda(k)]\psi_{0}(u+\alpha)A_{0}(k),~$$
 where $\Phi_{\alpha}(k,u)=\frac{\psi(i\frac{\partial}{\partial k},u+\alpha)\lambda(k)}{\Psi(i\frac{\partial}{\partial k},u+\alpha)\lambda(k)}$, and the term $\mathcal{I}(k,u)$ denotes the influence of the initial
condition. One can see that in above master equation advection and diffusion terms are coupled, and both depend on the reaction rate $\alpha$.
When the reaction rate $\alpha=0$, Eq.(\ref{Amaster}) reduces
to the master equation for CTRW in flows with the waiting time depending
on the preceding jump length in nonreactive system derived in Ref. [9].
Note also that for $\tau_{a}\rightarrow 0$, Eq.(\ref{i-(k,u)}) gives
\begin{equation}\label{i-(k,u)==}
i^{-}(k,u)\simeq[\psi(i\frac{\partial}{\partial
k},u+\alpha)\lambda(k)][i^{-}(k,u)+v_{k}i^{-'}_{k}(k,u)]+\psi_{0}(u+\alpha)A_{0}(k).
\end{equation}
Substituting (\ref{i-(k,u)=}) into (\ref{i-(k,u)==}), we can also find Eq.(14).

Analogously we shall now study the transport for the
B-particles in $A\rightarrow B$ reaction under energy-dependent anomalous diffusion in fluid fields. Let $B(x,t)$ be the PDF of B particle being in point
$x$ at time $t$, $j^{+}(x,t)$ be the gain flux and $j^{-}(x,t)$ be
the loss flux of particles B at site $x$ at $t$. Noting that
B-particle that is at (or leaves)  site $x$ at time $t$ either has
come there as a B-particle at some prior time or was converted from
an A-particle that either was on site $x$ from the very beginning or
arrived there later at $t'>0$ while keeping the same energy, and still stays (or just leaves)
the site $x$ at time $t$, we give the following balance
equations:
\begin{eqnarray}\label{B(x,t)}
B(x,t)&=&
\int_{-\infty}^{+\infty}dx'\int_{0}^{t}j^{-}(x',t')\lambda(x-x'-\tau_{a}v(x'))\Psi(|x-x'-\tau_{a}v(x')|,t-t')dt'
\nonumber\\
&&+\int_{-\infty}^{+\infty}dx')\int_{0}^{t}i^{-}(x',t')\lambda(x-x'-\tau_{a}v(x'))\Psi(|x-x'-\tau_{a}v(x')|,t-t')\nonumber\\
&&\times(1-e^{-\alpha(t-t')})dt'+A_{0}(x)\Psi_{0}(t)(1-e^{-\alpha t}),
\end{eqnarray}
and
\begin{eqnarray}\label{j-(x,t)}
j^{-}(x,t)&=&
\int_{-\infty}^{+\infty}dx'\int_{0}^{t}j^{-}(x',t')\lambda(x-x'-\tau_{a}v(x'))\psi(|x-x'-\tau_{a}v(x')|,t-t')dt'\nonumber\\
&&+\int_{-\infty}^{+\infty}dx'\int_{0}^{t}i^{-}(x',t')\lambda(x-x'-\tau_{a}v(x'))\psi(|x-x'-\tau_{a}v(x')|,t-t')\nonumber\\
&&\times(1-e^{-\alpha(t-t')})dt'+A_{0}(x)\psi_{0}(t)(1-e^{-\alpha t}),
\end{eqnarray}
where the initial condition $B_{0}(x)=0$ was used. Laplace
$x\rightarrow k$ and Fourier $t\rightarrow u$ transforming of
the two equations (\ref{B(x,t)}) and (\ref{j-(x,t)}) yields:
\begin{eqnarray}\label{B(k,u)}
B(k,u)&=&[\Psi(i\frac{\partial}{\partial k},u)\lambda(k)]\int_{-\infty}^{+\infty}j^{-}(x',u)e^{-k(x'+\tau_{a}v(x'))}dx'+[\Psi(i\frac{\partial}{\partial k},u)\lambda(k)-\Psi(i\frac{\partial}{\partial k},u+\alpha)\lambda(k)]\nonumber\\
&&\times\int_{-\infty}^{+\infty}i^{-}(x',u)e^{-k(x'+\tau_{a}v(x'))}dx'+A_{0}(k)[\Psi_{0}(u)-\Psi_{0}(u+\alpha)],
\end{eqnarray}
and
\begin{eqnarray}\label{j-(k,u)}
j^{-}(k,u)&=&[\psi(i\frac{\partial}{\partial k},u)\lambda(k)]\int_{-\infty}^{+\infty}j^{-}(x',u)e^{-k(x'+\tau_{a}v(x'))}dx'+[\psi(i\frac{\partial}{\partial k},u)\lambda(k)-\psi(i\frac{\partial}{\partial k},u+\alpha)\lambda(k)]\nonumber\\
&&\times\int_{-\infty}^{+\infty}i^{-}(x',u)e^{-k(x'+\tau_{a}v(x'))}dx'+A_{0}(k)[\psi_{0}(u)-\psi_{0}(u+\alpha)].
\end{eqnarray}
Here,  $B(k,u)$ and $j^{-}(k,u)$ are the Fourier-Laplace
transforms of $B(x,t)$ and $j^{-}(x,t)$, respectively.

Comparing (\ref{A(k,u)}), (\ref{i-(k,u)}), (\ref{B(k,u)}) and (\ref{j-(k,u)}), one has
\begin{eqnarray}\label{Brelation}
\frac{A(k,u)+B(k,u)-A_{0}(k)\Psi_{0}(u)}{\Psi(i\frac{\partial}{\partial k},u)\lambda(k)}=\frac{i^{-}(k,u)+j^{-}(k,u)-A_{0}(k)\psi_{0}(u)}{\psi(i\frac{\partial}{\partial k},u)\lambda(k)},
\end{eqnarray}
from which we find
\begin{equation}\label{j-(k,u)=}
j^{-}(k,u)=A_{0}(k)\psi_{0}(u)-i^{-}(k,u)+\Phi_{0}(k,u)[A(k,u)+B(k,u)-A_{0}(k)\Psi_{0}(u)].
\end{equation}
To get the master equation for $B(k,u)$, we consider the third balance equation$^{[12]}$:
\begin{equation}\label{B(x,t)otherbalance}
\frac{\partial B(x,t)}{\partial t}=j^{+}(x,t)-j^{-}(x,t)+\alpha
A(x,t).
\end{equation}
Here, $j^{+}(x,t)$ is the gain flux which can be represented by the
loss flux$^{[21]}$
\begin{equation}
j^{+}(x,t)=
 \int_{-\infty}^{+\infty}j^{-}(x',t)\lambda(x-x'-\tau_{a}v(x'))dx'.
 \end{equation}
Transforming $(x,t)\rightarrow (k,u)$  of (\ref{B(x,t)otherbalance}) yields
\begin{eqnarray}\label{B(k,u)otherbalance}
uB(k,u)=
\lambda(k)\int_{-\infty}^{+\infty}j^{-}(x',u)e^{-k(x'+\tau_{a}v(x'))}dx'-j^{-}(k,u)+\alpha A(k,u),
\end{eqnarray}
Assuming $v(x)=\omega x$, in the limit $\tau_{a}\rightarrow 0$, we find
\begin{eqnarray}\label{B(k,u)otherbalance}
uB(k,u)=
(\lambda(k)-1)j^{-}(k,u)+v_k\lambda(k)j^{-'}_{k}(k,u)+\alpha A(k,u),
\end{eqnarray}
 we substitute Eq.(\ref{j-(k,u)=}) into (\ref{B(k,u)otherbalance}),  and finally obtain the generalized master equation
in Fourier-Laplace space for the distribution of B-particles  in $A\rightarrow B$ reaction-anomalous diffusion on moving flows with waiting time depending the preceding jump length:
\begin{equation} \label{Bmaster}
\mathcal{D}_{1}(k,u)B(k,u)+\mathcal{D}_{2}(k,u)A(k,u)=\mathcal{V}_{1}(k,u)B_{k}^{'}(k,u)+\mathcal{V}_{2}(k,u)A_{k}^{'}(k,u)(k,u)+\mathcal{I}(k,u)
\end{equation}
where
$$\mathcal{D}_{1}(k,u)=u-(\lambda(k)-1)\Phi_{0}(k,u)-\lambda(k)v_{k}\Phi_{0}(k,u)^{'}_{k},~~~~~~~~~~~~~~~~~~~~~~~~~~~~~~~~~~~~~~~~~~~$$
$$\mathcal{D}_{2}(k,u)=(1-\lambda(k))[\Phi_{0}(k,u)-\Phi_{\alpha}(k,u)]+\lambda(k)v_{k}[\Phi_{0}(k,u)^{'}_{k}-\Phi_{\alpha}(k,u)^{'}_{k}]-\alpha,~~~~~~~~~$$
$$\mathcal{V}_{1}(k,u)=\lambda(k)v_{k}\Phi_{0}(k,u),~~~~~~~~~~~~~~~~~~~~~~~~~~~~~~~~~~~~~~~~~~~~~~~~~~~~~~~~~~~~~~~~~~~~~~~~~~~~~~$$
$$\mathcal{V}_{2}(k,u)=\lambda(k)v_{k}[\Phi_{0}(k,u)-\Phi_{\alpha}(k,u)],~~~~~~~~~~~~~~~~~~~~~~~~~~~~~~~~~~~~~~~~~~~~~~~~~~~~~~~~~~~~~$$
$$~\mathcal{I}(k,u)=[\lambda(k)-1]A_{0}(k)\{[\psi_{0}(u)-\psi_{0}(u+\alpha)]-[\Phi_{0}(k,u)\Psi_{0}(u)-\Phi_{\alpha}(k,u)\Psi_{0}(u+\alpha)]\}$$
$$~~~~~~~~~~~~+\lambda(k)v_{k}\{A^{'}_{0}(k)[\psi_{0}(u)-\psi_{0}(u+\alpha)]-A^{'}_{0}(k)[\Phi_{0}(k,u)\Psi_{0}(u)-\Phi_{\alpha}(k,u)\Psi_{0}(u+\alpha)]$$
$$~~~~~~~~~~~~~~~~~~-A_{0}(k)[\Phi_{0}(k,u)^{'}_{k}\Psi_{0}(u)-\Phi_{\alpha}(k,u)^{'}_{k}\Psi_{0}(u+\alpha)]\}~~~~~~~~~~~~~~~~~~~~~~~~~~~~~~~~~~~~~~~~~~~~~~~~~~$$
where the term $\mathcal{I}(k,u)$ denotes the influence of the initial
condition. Eq.(\ref{Bmaster}) shows the complex coupled relations among diffusion, advection and reaction terms. The generalized ADRE$'$s  for B-particles can be derived from  Eq.(\ref{Bmaster})  by carrying out the macroscopic limit
and inverting the Fourier-Laplace transform.

\section{ Examples and generalized ADRE$'$s }

We now turn to apply the master equations (\ref{Amaster}) and (\ref{Bmaster}) to derive the corresponding
 ADRE$^{'}$s for A and B particles when the jump lengths
obey Gaussian distribution or l$\acute{e}$vy flight and the waiting
time PDF  is quadratic dependent on the preceding jump length, i.e.,
\begin{equation}
\psi(|y|,t)=\delta(t-\theta y^{2}),
\end{equation}
where the length-dependent parameter $\theta>0$.

We first consider the case for Gaussian jump length PDF
\begin{equation}
\lambda(y)=\frac{1}{\sqrt{2\pi}\sigma}e^{-\frac{y^{2}}{2\sigma^{2}}}.
\end{equation}
 Then, the corresponding Laplace and Fourier transforms of
$\psi(|y|,t), \lambda(y)$ become
\begin{equation}
\psi(|y|,u)=e^{-\theta y^2 u}\sim 1-\theta y^2 u,
\end{equation}
\begin{equation}
\lambda(k)=e^{-\frac{\sigma^{2}k^{2}}{2}}\sim 1-\frac{\sigma^{2}k^{2}}{2},
\end{equation}
from which we find
\begin{equation}
\psi(i\frac{\partial}{\partial
k},u)\lambda(k)=1-\frac{\sigma^{2}k^{2}}{2}-\theta\mu\sigma^{2},
\end{equation}
\begin{equation}
\Psi(i\frac{\partial}{\partial k},u)\lambda(k)=\theta\sigma^{2}.
\end{equation}
Assuming $A_{0}(x)=\delta(x)$, substituting Eq.(31) and (32) and
the initial condition $\psi_{0}(u)=1$, $\Psi_{0}(u)=0$ into Eq.(\ref{Amaster}),
in the limit of $\tau_{a}\rightarrow 0$ and $\sigma\rightarrow 0$ we
obtain
\begin{equation}
(\frac{k^{2}}{2\theta}+ u+\alpha)A(k,u)=\frac{A}{\theta}\omega k\rho'_{k}(k,u)+1
\end{equation}
 Inverting
Eq.(33) to the space-time domain $k\rightarrow x$ and $s\rightarrow
t$, we then get the generalized ADRE for A-particles  with energy-dependent parameter:
\begin{equation}\label{ADREA}
\hat{T}_{t}(1,\alpha) A(x,t)+\frac{A}{\theta}\frac{\partial(v(x)A(x,t))}{\partial x}=\frac{1}{2\theta}\frac{\partial^{2}A(x,t)}{\partial x^{2}}
\end{equation}
with the initial condition $A_{0}(x)=\delta(x)$.
 Here, The integral
operator $\hat{T}_{t}(1-\beta,\alpha)f = \tau^{\beta}\Gamma
(1-\beta)\int_{0}^{t}\Phi_(\alpha)(t-t')f(t')dt' $ corresponds in
time domain to
\begin{eqnarray}
\hat{T}_{t}(1-\beta,\alpha)f&=&\frac{d
}{dt}\int_{0}^{t}\frac{e^{-\alpha(t-t')}}{(t-t')^{1-\beta}}f(t')dt'+\alpha\int_{0}^{t}\frac{e^{-\alpha(t-t')}}{(t-t')^{1-\beta}}f(t')dt'
\end{eqnarray}
and becomes a fractional derivative when $\alpha=0$ $^{[12]}$.
One can see that in the generalized ADRE (35) the diffusion
operate depend on reaction rate $\alpha$, and the advection and diffusion coefficients include the length-dependent parameter $\theta$.

If we substitute Eq.(31) and (32) and
the initial condition $B_{0}(x)=0$, $\psi_{0}(u)=1$, $\Psi_{0}(u)=0$ into Eq.(\ref{Bmaster}),
in the limit of $\tau_{a}\rightarrow 0$ and $\sigma\rightarrow 0$, we then
find
\begin{equation}
(\frac{k^{2}}{2\theta}+ u)B(k,u)=\frac{A}{\theta}\omega k B'_{k}(k,u)+\alpha A(k,u)
\end{equation}
 Inverting
Eq.(36) to the space-time domain , we obtain the generalized ADRE for B-particles:
\begin{equation}\label{ADREB}
\frac{\partial B(x,t)}{\partial t}+\frac{A}{\theta}\frac{\partial(v(x)B(x,t))}{\partial x}=\frac{1}{2\theta}\frac{\partial^{2}B(x,t)}{\partial x^{2}}+\alpha A(x,t)
\end{equation}
with the initial condition $B_{0}(x)=0$. Note that the advection and diffusion coefficients depend on  the length-dependent parameter $\theta$.

Secondly,  we choose a L$\acute{e}$vy distribution for the jump
length, i.e.,
\begin{equation}
\lambda(k)=e^{-\frac{1}{2}\sigma^{\beta}k^{\beta}}\sim 1-\frac{1}{2}\sigma^{\beta}k^{\beta}
\end{equation}
with $1<\beta\leq2$. Thus, for the length-dependent waiting time PDF
$$\psi(|y|,t)=\delta(t-\theta y^{2}),\theta>0,$$
we have
\begin{equation}
\psi(i\frac{\partial}{\partial
k},u)\lambda(k)=1-\frac{1}{2}\sigma^{\beta}k^{\beta}-\frac{1}{2}\theta
u \sigma^{\beta}\beta (\beta-1)k^{\beta-2},
\end{equation}
\begin{equation}
\Psi(i\frac{\partial}{\partial k},u)\lambda(k)=\frac{1}{2}\theta
\sigma^{\beta}\beta (\beta-1)k^{\beta-2}.
\end{equation}
For the initial condition $\rho_{0}(x)=\delta(x)$, in the limit of
small $\tau_{a}$ and $\sigma$, Eq.(\ref{Amaster}) then becomes
\begin{equation}
(\frac{1}{\theta\beta(\beta-1)}k^{2}+
u+\alpha)A(k,u)=A\frac{2}{\theta\beta(\beta-1)}\omega
k^{3-\beta}A'_{k}(k,u)+1,
\end{equation}
where $A=\lim_{\tau_{a}\rightarrow 0,\sigma\rightarrow
0}\frac{\tau_{a}}{\sigma^{\beta}}$ is kept finite. The inverse
Fourier-Laplace transform of Eq.(30) leads to the generalized
fractional ADRE for A-particles:
\begin{equation}\label{FADREA}
\hat{T}_{t}(1,\alpha) A(x,t)-\frac{2Ai^{\beta}}{\theta\beta(\beta-1)}D_{x}^{3-\beta}(v(x)A(x,t))=\frac{1}{\theta\beta(\beta-1)}\frac{\partial^{2}A(x,t)}{\partial
x^{2}}
\end{equation}
with the initial condition $\rho_{0}(x)=\delta(x)$. Here, the
operator $D_{x}^{3-\beta}$, is the fractional derivative of the
Riemann-Liouville type,$^{[22]}$ equal in Fourier $x\rightarrow
k$ space to $(ik)^{3-\beta}$. In (42)  the
advection and diffusion coefficients both involve the
length-dependent parameter $\theta$. Note also that this generalized
fractional equation (42) reduces to Eq.(34) when $\beta=2$.

Analogously for  $B_{0}(x)=0$, in the limit of
small $\tau_{a}$ and $\sigma$, Eq.(\ref{Bmaster}) becomes
\begin{equation}
(\frac{1}{\theta\beta(\beta-1)}k^{2}+
u)B(k,u)=A\frac{2}{\theta\beta(\beta-1)}\omega
k^{3-\beta}B'_{k}(k,u)+\alpha A(k,u),
\end{equation}
Inversing
 Eq.(43), we obtain the generalized
fractional ADRE for B-particles  with energy-dependent parameter $\theta$:
\begin{equation}\label{FADREB}
\frac{\partial B(x,t)}{\partial t}-\frac{2Ai^{\beta}}{\theta\beta(\beta-1)}D_{x}^{3-\beta}(v(x)B(x,t))=\frac{1}{\theta\beta(\beta-1)}\frac{\partial^{2}B(x,t)}{\partial
x^{2}}+\alpha A(x,t).
\end{equation}
If $\beta=2$, this generalized
fractional equation then reduces to Eq.(37).

Finally, assume that
$C(x,t)$ is the sum of $A(x,t)$ and $B(x,t)$, and combine (\ref{ADREA})
with (\ref{ADREB}), and we have
\begin{equation}\label{GADE}
\frac{\partial C(x,t)}{\partial t}+\frac{A}{\theta}\frac{\partial(v(x)C(x,t))}{\partial x}=\frac{1}{2\theta}\frac{\partial^{2}C(x,t)}{\partial x^{2}}.
\end{equation}
From (\ref{FADREA})
and (\ref{FADREB}), one has
\begin{equation}\label{GFADE}
\frac{\partial C(x,t)}{\partial
t}-\frac{2Ai^{\beta}}{\theta\beta(\beta-1)}D_{x}^{3-\beta}(v(x)C(x,t))=\frac{1}{\theta\beta(\beta-1)}\frac{\partial^{2}C(x,t)}{\partial
x^{2}}.
\end{equation}
Note that Eq.(\ref{GADE}) and (\ref{GFADE}) are consistent with  the generalized advection-dispersion equations with waiting time depending on the preceding jump length  derived in Ref. [9] recently.
Note also that in above two equation there are no reaction terms except diffusion terms. It is because that the $A\rightarrow B$ reaction we discuss here does not
change the sum of the particles in the reactive system.

\section{Conclusions}
To sum up, in this paper we derive the master equations (\ref{Amaster}) and (\ref{Bmaster}) for the PDF
of A and B particles in reaction
$A\rightarrow B$  under
anomalous diffusion in linear flows with random waiting time depending on the preceding jump length based on the CTRW model. As examples, we obtain four generalized
ADRE's (\ref{ADREA}), (\ref{ADREB}), (\ref{FADREA}) and (\ref{FADREB}) for  the probability density function of reactive particles by applying the derived master equations, and show the energy-dependent characteristics of the particles in fluid reactive anomalous transport. There are problems such as the energy-dependent
behaviors for more complex reaction under anomalous diffusion on moving fluid are still unknown.

\vspace*{2mm}

%%%% �ο������Ű���ʽ��

\end{CJK*}
\end{document}